\documentclass[aps,prc,preprint,groupedaddress]{revtex4}

\usepackage{graphicx}
\usepackage{amssymb}

\begin{document}


\title{Two models with rescattering for high energy heavy ion collisions}


\author{H. B{\o}ggild}
\email[]{boggild@nbi.dk}
\author{Ole Hansen}
\email[]{ohansen@nbi.dk}
\affiliation{Univ. of Copenhagen, The Niels Bohr Institute,
Copenhagen, Denmark}
\author{T. J. Humanic}
\email[]{humanic@mps.ohio-state.edu}
\thanks{The authors wish to acknowledge financial
support from the U.S. National Science Foundation under grant
PHY-0355007 and from the Danish SNF for travel expenses.}
\affiliation{Department of Physics, The Ohio State University,
Columbus, Ohio, USA}


\date{\today}

\begin{abstract}
The effects of hadronic rescattering in RHIC-energy Au+Au collisions
are studied using two very different models to describe the early
stages of the collision. One model is based on a hadronic thermal
picture and the other on a superposition of parton-parton
collisions. Operationally, the output hadrons from each of these
models are used as input to a hadronic rescattering calculation. The
results of the rescattering calculations from each model are then
compared with rapidity and transverse momentum distributions from
the RHIC BRAHMS experiment. It is found that in spite of the
different points of view of the two models of the initial stage,
after rescattering the observed differences between the models are
mostly ``washed out" and both models give observables that agree
reasonably well with each other and experiment.
\end{abstract}

\pacs{25.75.Dw, 25.75.Gz, 25.75.Ld}

\maketitle


\section{Introduction}
The main goal of studying relativistic heavy ion collisions at the
RHIC accelerator is to obtain information about the early stage of
the collision when matter is at its most hot and dense state. Since
experimentally one detects hadrons  which have undergone final-state
rescattering before decoupling from the collision, it is of interest
to use model calculations to seek to disentangle the hadronic
rescattering effect that tend to wash out the information about the
early state of matter in which we are most interested. Such a
rescattering calculation was carried out for RHIC collisions
assuming a simple thermal-like model to describe the early stage of
the collision \cite{Humanic:2002b,Humanic:2006a,Humanic:2006b}, but
since this thermal model was so simple, it proved difficult to
interpret the results for the initial stage. The present work
improves on the previous study in two ways: 1) a more elaborate
thermal-like model is used for the initial stage, and 2) a second
initial-stage model based on a superposition of parton-parton
collisions is also included in the study. The advantage of 1) is
clearly to make the interpretation of the initial-stage results
easier, and the advantage of 2) is to compare the results of the
thermal-like model with a model from a very different point of view,
i.e. partonic, to see if after rescattering identifiable features of
the different initial-stage models survive. We thus hope to address,
at least for these two models, to what extent rescattering washes
out the information about the initial stage of the collision. Our
comparisons will be made with the hadronic observables rapidity and
transverse momentum distributions, and these in turn will be
compared with those extracted from the RHIC BRAHMS experiment
\cite{Peter 2003,Djamel 2004}, as well as other RHIC experiments.

Sections II, III, and IV describe the thermal model, parton model,
and rescattering calculation method, respectively. Sections V and VI
give results of coupling the rescattering calculation with the
thermal and parton models, respectively, and of comparisons with the
BRAHMS experiment. Section VII presents a discussion of the results.

\section{The thermal-like model}
\subsection{Overview}
The thermal-like model that we use builds on the Bjorken picture of
a high energy heavy ion reaction \cite{Bjorken:1983a}. The two heavy
ions pass through one another in a central collision, whereby both
nuclei become highly excited and a color field of high energy
density is created in the space between the two ions after the
collision. Particles are produced, in part from the two original
nuclei with a net-baryon number to insure baryon conservation and in
part from the region between the two ions, a region with a near
vanishing net baryon content. It is assumed here that the produced
particles may be described as originating from three thermal source
centers, corresponding to the two heavy ions and the energy field in
between. The source centers are extended in rapidity space, in the
present model each distributed over a Gaussian shaped rapidity
probability density. All three source centers are assumed to have
the same temperature, $T$, but may contribute different numbers of
particles. The created particles are assumed to have energy
distributions that follow the Boltzmann distribution. The thermal
model presented below thus for each type of particle creates three
pools of four-momentum vectors, each pool distributed in rapidity
around a source center. The differential distributions are then
created by a weighted sum over contributions from the three source
centers, where the weights, $N_{C}$, are the number of four-momentum
vectors (particles) from each center.

It is well documented from previous work \cite{Johanna etc}, that
particle ratios, in principle integrated over the entire phase
space, are described very well by thermal statistical ensembles, but
thermal models that give insight into differential distributions in
rapidity, transverse mass or momentum are scarce. To allow for the
non-thermal phenomenon of flow, observed by experiments, the
thermally produced particles are here allowed to rescatter as they
emerge.

The purpose of the exercise presented here, is to show such
differential distributions, compare them with data and obtain some
understanding as to the extent such a description can reproduce the
main features of the observations.

\subsection{Structure of the model}
For a given particle (mass=$m$) and a given source-center
(rapidity=$y_{C}$) the four-momentum vectors are created by four
subsequent Monte Carlo routines. The first one chooses the rapidity
$y_{G}$ of the local source as a deviation from $y_{C}$ via a
Gaussian probability density distribution,
\begin{equation}
G(y_{G},y_{C}) = \frac{1}{\sigma_{C}\sqrt{2\pi}}
\exp{(-(y_{G}-y_{C})^{2}/2\sigma_{C}^{2})},
\end{equation}
where $\sigma_{C}$ characterizes the width of the Gaussian
distribution corresponding to the selected source-center. The four
momentum is generated in the local source reference system by
choosing the polar and azimuthal angles, $\theta_{B}$ and
$\phi_{B}$, such that the polar angle is taken from a constant
distribution in $\cos(\theta_{B})$ for $0<\theta_{B}<\pi$, and
$\phi_{B}$ is evenly distributed from 0 to $2\pi$. The energy of the
particle, E$_{B}$, is finally chosen according to a Boltzmann
prescription,
\begin{equation}
B(E_{B},T)=\frac{E_{B}\sqrt{E_{B}^{2}-m^{2}}}{Tm^{2}K_{2}(m/T)}
\exp{(-E_{B}/T)}.
\end{equation}
The temperature is denoted $T$ and is a global parameter used for
all particle types, source-centers, and local sources. The $K_{2}$
is a modified Bessel function \cite{Assimof}. The four-momentum
components ($E_{B}$, $p_{x}$, $p_{y}$, $p_{z}$) following from the
above Monte Carlo choices are then Lorentz transformed to the
laboratory system and the process started over again with a new
choice of $y_{G}$.

Each particle thus originates with its four-momentum from its own
local source reference system. If the width $\sigma_{C}$=0, the
particles from source-center C, will represent particles from a
spherically symmetric Boltzmann source in the source-center
reference system. A certain number of particles, $N_{m,C}$ are
generated from each source-center, respectively, and the collection
of four-vectors then constitutes the model  data for particle
species $m$. The multiplicity density distribution is then obtained
as the sum of three source-center contributions
\begin{equation}
dn/dy_{m}(y)=N_{m0}F_{m0}(y)+N_{m-}F_{m-}(y)+N_{m+}F_{m+}(y),
\end{equation}
where $y$ is the laboratory rapidity and the F-functions follow from
the computation as described above.

\subsection{Parameters of the model}
For a particle of mass $m$, the model has 10 parameters: the
temperature $T$, the rapidities of the three source-centers $y_{+}$,
$y_{0}$ and $y_{-}$, the widths of the three Gaussian distributions
$\sigma_{+}$, $\sigma_{0}$ and $\sigma_{-}$, and the number of
particles from each source-center $N_{+}$, $N_{0}$ and $N_{-}$. In
this report only symmetric collisions, A+A, are considered, which
impose four restrictions on the parameters,
\begin{equation}
y_{0}=0
\end{equation}
\begin{equation}
y_{+}=-y_{-}
\end{equation}
\begin{equation}
\sigma_{+}=\sigma_{-}
\end{equation}
\begin{equation}
N_{m+}=N_{m-}
\end{equation}
The temperature for the following is set at 200 MeV, but later in
the report it is increased to 270 MeV. The $N_{mC}$ values (number
of particles or multiplicities) are chosen for each $(m,C)$
combination. The ratio $N_{+}/N_{0}$ has a decisive influence on the
shape of the predicted rapidity density distribution, $dn/dy$, and
on the slope of the transverse spectra. The same holds true for the
$y_{+}$, the $\sigma_{+}$ and $\sigma_{0}$. The parameters were
determined by asking for a reasonable agreement with the proton
$dn/dy$ distribution for Au+Au at $\sqrt{s}$=200 GeV per nucleon as
measured by the BRAHMS collaboration \cite{Peter 2003}, in  the
expectation that the $y_{+}$, $\sigma_{+}$ and $\sigma_{0}$,
determining the Gaussian distributions, would be useful for all the
particle species considered, an expectation that was fulfilled.

The BRAHMS proton data show that $y_{+}$ is larger than 3.0, but
they do not fix the value, because of the limited rapidity coverage
of the experiment. The value used here of $y_{+}$=3.5 is a
reasonable value, in particular when the measured distribution of
net-protons is also considered, but the best value might be larger.
The shape of a predicted $dn/dy$ distribution is changed moderately,
but not drastically, by introducing the rescattering routines (see
later), a fact that cuts down the computing time for obtaining the
needed parameters, because the fitting could be done quite reliably
without the rescattering. In fact the computing time with
rescattering would have rendered the fitting process impractical.

\begin{figure}[ht!]
\begin{center}
\includegraphics[height=10cm]{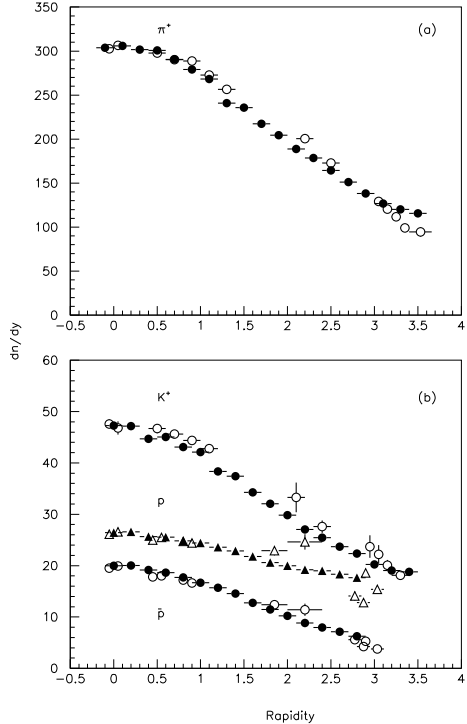} \caption{The rapidity
density $dn/dy$ plotted against center of mass rapidity for
$\pi^{+}$ (top), $K^{+}$, $p$, and $\bar{p}$ (bottom). Open symbols
designate data from BRAHMS \cite{Peter 2003,Djamel 2004} while the
black symbols are used for thermal model data, $T=200$ MeV.}
\label{fig1}
\end{center}
\end{figure}

The actual fitting was made by varying the Gaussian sigmas and the
proton $N_{+}$ and $N_{0}$ in a trial-and-error way. The final
values, kept constant for the remaining use of the model here, are
given in Table I. They do not necessarily represent a best fit
(which was never sought after), but they do represent a reasonable
fit. A comparison between the model and the BRAHMS proton data
\cite{Peter 2003} for dn/dy is shown in the bottom part of Figure
\ref{fig1}. For all other particles ($\bar{p}$, $\pi^{+}$ and
$K^{+}$) only the corresponding $N_{+}$ and $N_{0}$ values were
adjusted to give the measured ratio between $dn/dy$ at $y=0$ and at
the highest value of $y$ at which there were data \cite{Djamel 2004}
for the particle type in question. The $N_{+}/N_{0}$-values obtained
are shown in Table I, and comparisons to the BRAHMS data in Figure
\ref{fig1}.
Transverse spectra may be constructed from the model data by
selecting a rapidity interval and sampling the number of particles
as a function of $m_{t}$ or $p_{t}$. For a given species the
spectrum is again made as a weighted sum of contributions from the
three source centers,
\begin{equation}
Ed^{3}n/dp^{3}=N_{m0}f_{m0}(m_{t})+N_{m+}f_{m+}(m_{t})+N_{m-}f_{m-}(m_{t}),
\end{equation}
where the f-functions indicate the spectral shape. The measured
transverse spectra were not used in the parameter search. The shapes
of the model spectra are more sensitive to the rescattering
routines, and a discussion of spectra and comparisons to
measurements are deferred to later in the paper. Suffice to say
here, the near exponential fall off of model $p_{t}$ and $m_{t}$
spectra with increasing transverse momentum is far too fast for
protons, antiprotons and $K^{+}$ as compared to the BRAHMS data, but
in fairly good agreement for $\pi^{+}$ spectra, all compared at
$y=0$ and at a rapidity near 3.10.
\begin{center}
\begin{table}
\caption{Fit parameters of the thermal model.}
\begin{tabular}{ccrr} \hline
Particle    & Quantity & Values & \\ \hline
all           & T (GeV) & 0.20  &   \\
all           & y$_{0}$, y$_{+}$ & 0.00 & 3.50 \\
all           & $\sigma_{0}, \sigma_{+}$ & 1.50 & 2.00 \\ \hline
p             & N$_{+}$/N$_{0}$ & 0.95 &   \\
$\bar{p}$ &                             & 0.18 &   \\
$\pi^{+}$ &                             & 0.40 &   \\
K$^{+}$  &                             & 0.45 &    \\ \hline
\end{tabular}
\end{table}
\end{center}

\section{The partonic jet model}
\subsection{Overview}
The ``partonic jet model" used in this study consists of two parts.
One is a simple model for pp, or rather nucleon-nucleon collisions,
each of which consists of a partonic collision leading to two jets
and  two ongoing wounded nucleons. The second part is a model for AA
collisions, which depending on the value of the impact parameter,
leads to a number of such binary partonic collisions with onward
moving wounded nucleons, allowed to re-interact. In both models
energy and momentum are approximately conserved by keeping track of
the energy used in each step.

\subsection{The pp Model}
\begin{figure}[ht!]
\begin{center}
\includegraphics[height=8cm]{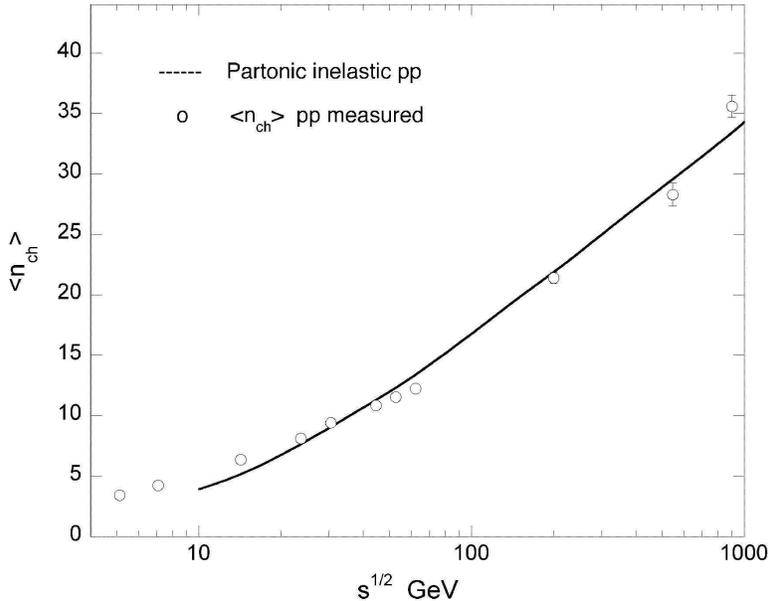} \caption{Mean charged
multiplicity in pp-collisions.} \label{fig2}
\end{center}
\end{figure}

In the present approach each nucleon-nucleon collision has a hard
scattering between two partons, leading to a system of two back to
back $e^{+}e^{-}$-like jets with multiplicities as a function of
center of mass (cms) energy and particle composition obtained  from
$e^{+}e^{-}$ data \cite{e-jets}. The collision is taking up a
certain fraction, x, of the energy of each of the incoming nucleons,
picked from an ``effective" structure function distribution. A very
simple probability distribution function is found adequate, i.e.
$dW/dx = 2*(1-x)$ , such that on average one third of the nucleon
incoming energy participates in the hard scattering. Since the two
parton energies are independent, the cms for the two-jet system is
not the overall cms, but different from event to event.The remaining
forward going nucleon systems are considered excited systems, each
of which fragments into one nucleon and one $e^{+}e^{-}$-like jet.
Here it is assumed that on average half the energy is taken by the
nucleon, i.e.  a flat x-spectrum is used for the nucleon energy.
Figure \ref{fig2}  shows how this model reproduces the mean charged
multiplicity in pp collisions in the energy range from 10 to 1000
GeV. \par In the jet fragmentation the multiplicity of each jet is
taken from a negative binomial distribution with the value of $k$
varying (decreasing) as a function of $\sqrt{s}$ and the
longitudinal x-distribution is subsequently determined by
one-dimensional longitudinal phase space.\par The transverse
momentum is generated  by a procedure taking into account the effect
of gluon bremsstralung. This is done by giving the mean $p_t$ two
components, one which is assumed constant $p_t^{out}$, and one
increasing linearly with $\sqrt{s}$, $p_t^{in}$. This is in good
qualitative agreement with $e^{+}e^{-}$ data, ref. \cite{inout}. We
find that adding (in quadrature) the transverse momenta from two
independent $m_{t}$ distributions with inverse slopes of 0.090
GeV/c$^{2}$ for $p_t^{out}$ and 0.180+($\sqrt{s}$-20)*0.001 for
$p_t^{in}$ gives a good agreement between the model and pp data, see
Figure \ref{fig4}. Note that the effective parton-parton scattering
angle distribution (see below) is also involved in generating the
$p_{t}$ distributions shown. For forward nucleons a $p_{t}$
distribution with a negative inverse slope of 0.175 GeV/c is used.
The next step is to obtain an effective parton-parton scattering
angle distribution for the two-jet system. In this work a function
from D. Perkins ( ref.\cite{Cambridge}) is used :
\begin{equation}
d\sigma/dcos\theta =
const*(3+cos^{2}\theta)^{3}/(1-cos^{2}\theta)^{2},
\end{equation}
\noindent An angle is found above a cut-off $p_{t,cut}$  following
the above probability distribution and the corresponding jet $p_{t}$
is calculated. Finally this $p_{t}$ is reduced by the cut-off value
: $p_{t,jet}=\sqrt{p_{t}^2-p_{t,cut}^2}$.  The used value of the jet
$p_{t}$-cutoff is $p_{t,cut}$= 0.3 GeV/c.

\begin{figure}[ht!]
\begin{center}
\includegraphics[height=7.0cm]{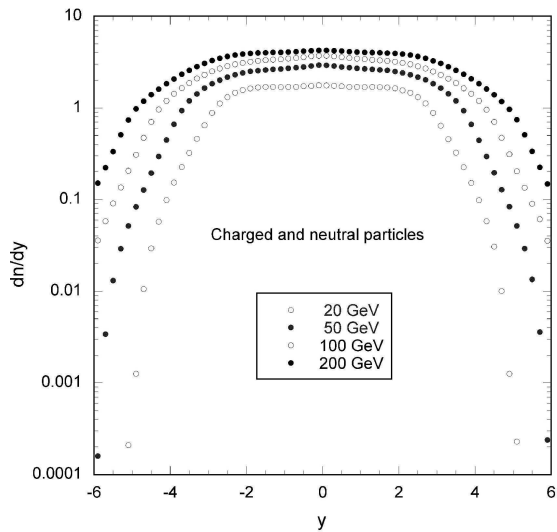} \caption{
Rapidity-distributions in pp collisions at cms energies of 20, 50,
100 and 200 GeV from the partonic model.} \label{fig3}
\end{center}
\end{figure}

It should be noted that the parton-parton sub system production and
fragmentation do not depend on the overall nucleon-nucleon (or AA)
collision energy. Figures \ref{fig3} and \ref{fig4} show the
rapidity and $p_{t}$ distributions obtained at different energies
and illustrates in the $p_{t}$ case that qualitative agreement  with
the data is obtained.

\begin{figure}[ht!]
\begin{center}
\includegraphics[height=7.0cm]{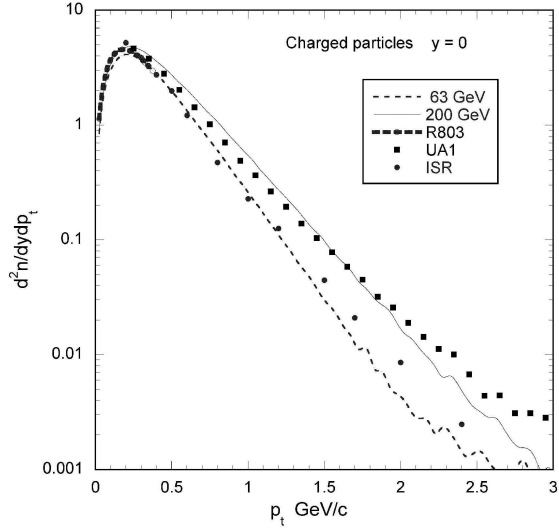} \caption{$p_{t}$
distributions from pp.The lower squares are from UA1 data at  $
\sqrt{s} = 200$ GeV and the upper from from the ISR at $\sqrt{s} =
50$ GeV. The dashed curves are partonic model results at 63 GeV
(lower) and 200 GeV (upper).} \label{fig4}
\end{center}
\end{figure}

\subsection{The AA Model}
The AA model is inspired by the work of Jackson and Boggild
\cite{JandB} and subsequent work \cite{JBblacklabel}. Each AA
collision, with specified impact parameter, involves a certain
number of binary collisions, $N_b$, and of participants, $N_p$. The
ratio of these numbers is the average number of collisions each
struck nucleon experiences, $N_c = N_b/N_p$. If we now let a $train$
of $N_c$ nucleons from one nucleus collide  with a similar $train$
from the other nucleus and do this $N_p/N_c$ times we get the right
number of binary collisions. In each $train-train$ collision the
procedure is the following using the above described parton inspired
pp model:

a) The two $e^{+}e^{-}$  like jets escape the collision and
fragment.

b) The forward wounded nucleon re-interacts with reduced energy.

c) At the end of the $train$ the wounded nucleon fragments as in the
pp model.

\noindent\\
In this way a collision of a $train$ of five against five nucleons
will produce 10 wounded nucleons with successively reduced energy
and 25 $e^{+}e^{-}$ like two-jet systems, in total leading to 50
(binary)+10(fragment) jets and ten nucleons.

\begin{figure}[ht!]
\begin{center}
\includegraphics[height=7.5cm]{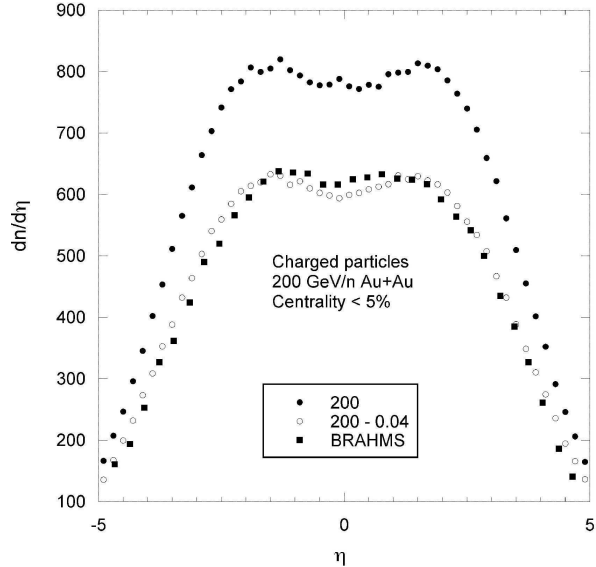}
\caption{Pseudo-rapidity distributions for central Au+Au collisions
at a cms energy of 200 GeV compared with data from Brahms
\cite{Brahms}. Full circles mark calculation at 200 GeV without pion
coalescence (see text subsect. 3.4), while open circles denote the
calculation including pion coalescence.} \label{fig5}
\end{center}
\end{figure}

Figure \ref{fig5} shows results of the model for central collisions
of gold on gold, i.e. the pseudo-rapidity distributions before and
after coalescence (see below) compared with BRAHMS data
\cite{Brahms}.

\subsection{A modification of the Model}
From Figure \ref{fig5} it is clear that the AA model leads to
charged particle multiplicities which are too high by 30-40\%. This
is not surprising since the model does not take shadowing
\cite{shadow} into account. Trying to model this by simply lowering
the multiplicity does not work because energy conservation forces
particles in the forward direction to take up the missing energy.
This leads to a forward peaked rapidity distribution. To lower the
multiplicity while maintaining energy-momentum conservation and the
shape of the rapidity distribution  a scheme of pion coalescing is
adopted; the procedure used is the following :
\par\noindent\\
a) The $Q^2_{inv}$ of all pion pairs is calculated, where \\
$Q^2_{inv}$ =
$(E_1-E_2)^2-(p_{x1}-p_{x2})^2-(p_{y1}-p_{y2})^2-(p_{z1}-p_{z2})^2$,
and $E$ designates the total energy of the pion.
\\
\noindent
b) Pion pairs with $-Q^2_{inv}$ lower than a certain cutoff are coalesced.\\
\par
\noindent It turns out that a cutoff at $-Q^2_{inv} = 0.04^{2}$
GeV$^{2}$ leads to a reasonable reduction of the charged particle
multiplicities and, at the same time, to a good agreement with
observed rapidity distributions as demonstrated for BRAHMS in Figure
\ref{fig5}. The assumption behind the above phenomenological
procedure is that nearby pions are for some time after creation
still overlapping and can act together. As further discussed below
the price paid is a hardening of the pion $p_{t}$ distribution where
the density is high.

The partonic model presented above in several respects resemble the
HIJING-model of Wong and Gyulassy \cite{hiji}.

\section{Hadronic rescattering calculation}
The rescattering model calculational method used is similar to that
employed in previous calculations for lower CERN Super Proton
Synchrotron (SPS) energies  and RHIC studies \cite{Humanic:2002b}.
Rescattering is simulated with a semi-classical Monte Carlo
calculation which assumes strong binary collisions between hadrons.
The Monte Carlo calculation is carried out in three stages: 1)
initialization and hadronization, 2) rescattering and freeze out,
and 3) calculation of experimental observables. Relativistic
kinematics is used throughout.

The hadronization model inputs momentum vectors from the thermal
model or partonic jet model both described above and employs simple
parameterizations to describe the initial space-time of the hadrons
similar to that used by Herrmann and Bertsch \cite{Herrmann:1995a}.
The initial space-time of the hadrons for $b=0$ fm (i.e. zero impact
parameter or central collisions) is parameterized as having
cylindrical symmetry with respect to the beam axis. The transverse
particle density dependence is assumed to be that of a projected
uniform sphere of radius equal to the projectile radius, $R$
($R={r_0}A^{1/3}$, where ${r_0}=1.12$ fm and $A$ is the atomic mass
number of the projectile). The initial transverse coordinates of a
given particle, i.e. $x_{had}$ and $y_{had}$, are thus determined
according to this distribution. The longitudinal particle
hadronization position ($z_{had}$) and time ($t_{had}$) are
determined by the relativistic equations \cite{Bjorken:1983a},
\begin{equation}
z_{had}=\tau_{had}\sinh{y_{i}}; \hspace{3mm}
t_{had}=\tau_{had}\cosh{y_{i}}
\end{equation}
\noindent where $y_{i}$ is the initial particle rapidity and
$\tau_{had}$ is the hadronization proper time. Thus the space-time
hadronization model has one free parameter to extract from
experiment: $\tau_{had}$. Although only pions, kaons, and nucleons
are input from the thermal model as the initial particle types for
the rescattering calculation, other types of hadrons can be produced
during rescattering. In all, the hadrons included in the calculation
are pions, kaons, nucleons, and lambdas ($\pi$, $K$, $N$, and
$\Lambda$), and the $\rho$, $\omega$, $\eta$, ${\eta}^{*}$, $\phi$,
$\Delta$, and $K^*$ resonances. For simplicity, the calculation is
isospin averaged (e.g. no distinction is made among a $\pi^{+}$,
$\pi^0$, and $\pi^{-}$).

The second stage in the calculation is rescattering which finishes
with the freeze out and decay of all particles. Starting from the
initial stage ($t=0$ fm/c), the positions of all particles are
allowed to evolve in time in small time steps ($\Delta t=0.1$ fm/c)
according to their initial momenta. At each time step each particle
is checked to see a) if it has hadronized ($t>t_{had}$), b) if it
decays, and c) if it is sufficiently close to another particle to
scatter with it. Isospin-averaged s-wave and p-wave cross sections
for meson scattering are obtained from Prakash et al.
\cite{Prakash:1993a}. The calculation is carried out to 100 fm/c,
although most of the rescattering finishes by about 50 fm/c. The
rescattering calculation is described in more detail elsewhere
\cite{Humanic:2006a,Humanic:1998a}. The validity of the numerical
methods used in the rescattering code have recently been studied and
verified\cite{Humanic:2006b}.

In the last stage of the calculation, the freeze-out and decay
momenta and space-times are used to produce observables such as
pion, kaon, and nucleon multiplicities and transverse momentum and
rapidity distributions. The values of the initial pion, kaon, and
nucleon multiplicities, temperature, and hadronization proper time
are all constrained to give observables which agree with available
measured hadronic observables. As a cross-check on this, the total
energy from the calculation is determined and compared with the RHIC
center of mass energy of $\sqrt{s}=200$ GeV to see that they are in
reasonable agreement. Particle multiplicities were estimated from
the charged hadron multiplicity measurements of the RHIC PHOBOS
experiment \cite{Back:2000a}. Calculations were carried out using
isospin-summed events containing at freezeout for central collisions
($b=0$ fm) about 5000 pions, 500 kaons, and 650 nucleons
($\Lambda$'s were decayed). The hadronization model parameter
$\tau_{had}$=1 fm/c was used. It is interesting to note that the
same value of $\tau_{had}$ was required in a previous rescattering
calculation to successfully describe results from SPS Pb+Pb
collisions \cite{Humanic:1998a}.

\section{Results from the thermal model with rescattering}
\subsection{$dn/dy$ with and without rescattering}
This subsection demonstrates the changes in the rapidity density
distributions caused by the rescattering. The rescattering routine
was run event by event and 20 events made up the total final event
pool, which was analyzed into $dn/dy$ and invariant cross section
distributions. The thermal model sometimes produces particles with
very large rapidities (e.g. $\mid y \mid \geq $10) in the forward
and backward directions. These particles do not have a counterpart
in a collision situation at any existing accelerator, so all
particles with $\mid y \mid \geq $ 6.5 were disregarded in the
rescattering calculation. $y=6.5$ is the beam rapidity at RHIC for
the data used in the subsequent comparisons (subsect. 5.3). Such
high rapidity particles constituted about 9\% of the four-vectors
generated by the thermal model and after their removal, the total
energy of the remaining particles in a $T=200$ MeV event was close
to the total energy in a $\sqrt{s}=200$ GeV Au+Au collision.
\begin{figure}[h!]
\begin{center}
\includegraphics[height=7.5cm]{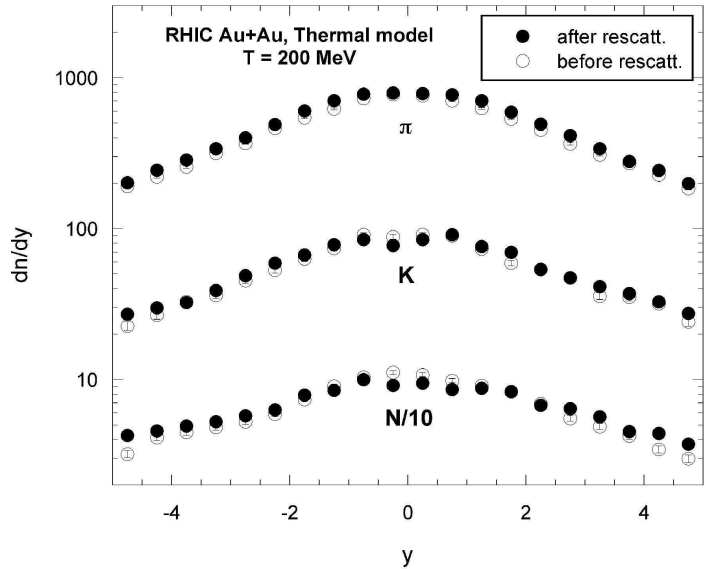} \caption{Rapidity
distributions from the thermal model with $T=200$ MeV; open circles
denote the results before rescattering and the black circles after.
$N/10$ in the figure stands for nucleon multiplicity divided by 10.}
\label{fig6}
\end{center}
\end{figure}
The resulting $dn/dy$ distributions at a temperature of $T=200$ MeV
for pions, kaons and nucleons are shown in Figure \ref{fig6}, where
open circles denote the distributions before rescattering and black
circles after rescattering. Figure \ref{fig7} shows results at
$T=270$ MeV. In all cases the influence of the rescattering is
finite and rather small. For nucleons the dn/dy with rescattering is
lower by less than 5\% at small $y$, than without rescattering, and
higher by a similar amount at high rapidities. Kaons are changed
less and in a similar way, and the pions are still less changed.
Also the changes in going from 200 MeV to 270 MeV are quite small.
\begin{figure}[h!]
\begin{center}
\includegraphics[height=7.5cm]{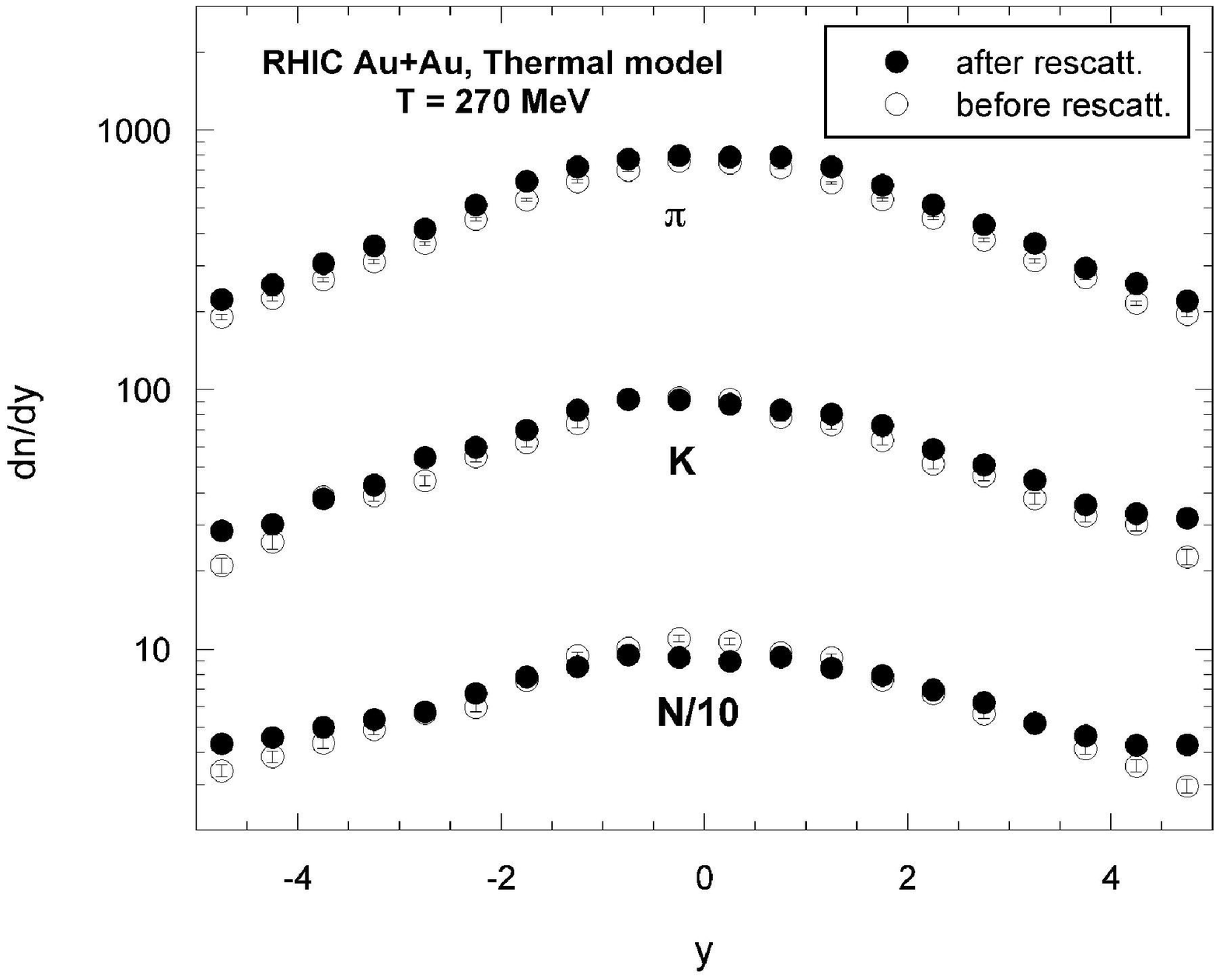} \caption{Rapidity
distributions from the thermal model with $T=270$ MeV; open circles
denote the results before rescattering and the black circles after.
$N/10$ in the figure stands for nucleon multipicity divided by 10.}
\label{fig7}
\end{center}
\end{figure}

\subsection{Invariant multiplicities with and without rescattering}
The invariant multiplicities are analyzed as functions of either
transverse momentum, $p_{t}$, or transverse mass,
$m_{t}=\sqrt{p_{t}^{2}+m^{2}}$:
\begin{equation}
E\frac{d^{3}n}{dp^{3}}=\frac{d^{2}n}{2\pi m_{t}dm_{t}dy}=
\frac{d^{2}n}{2\pi p_{t}dp_{t}dy},
\end{equation}
\noindent where the multiplicity is denoted by $n$. In all cases
shown the rapidity width in the model analysis is $\Delta y=1.0$,
while the $m_{t}$ interval was 0.2 GeV.

For an isotropic Boltzmann source the shape of the invariant
$m_{t}$-spectrum reflects the temperature $T$ as
\begin{equation}
\frac{1}{2\pi m_{t}} \frac{d^{2}n}{dm_{t}
dy}=N_{Bol}m_{t}\cosh(y-y_{C})
\exp{(-\frac{m_{t}}{T/\cosh(y-y_{C})})},
\end{equation}
\noindent where $y_{C}$ is the rapidity of the source and $N_{Bol}$
the usual Boltzmann normalization (see also Eq. (2). The present
thermal model represents the sum over many single particle sources,
centered around three source centers, and will not show spectra
following Eq. (12), nor the $1/\cosh(y-y_{C})$ dependence of the
apparent temperature. None the less, the $m_{t}$-spectra are nearly
exponential as in Eq. (12), and they exhibit an inverse slope that
varies with rapidity in a way that also depends on particle mass, a
feature that is different from the single spherical Boltzmann
source, where the mass only enters explicitly in the normalization.

\begin{figure}[h!]
\begin{center}
\includegraphics[height=8cm]{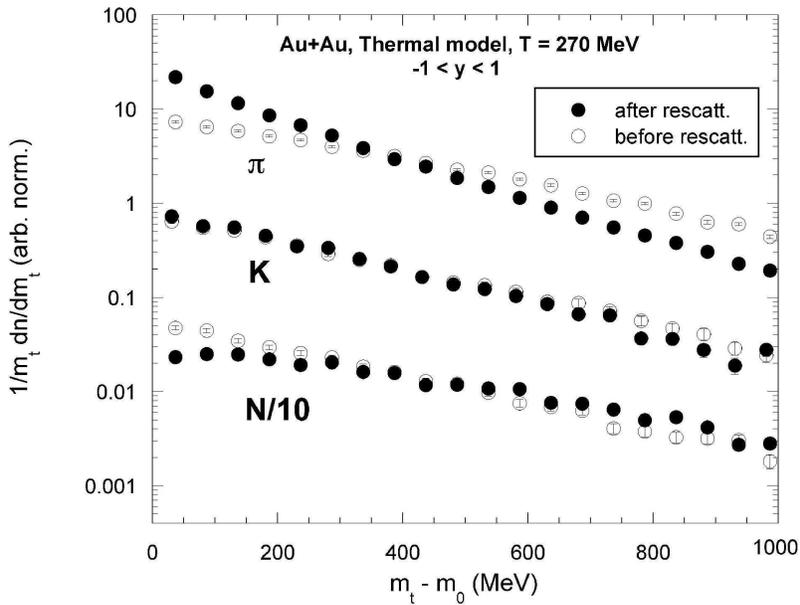} \caption{Transverse mass
spectra for $T=270$ MeV collected for rapidities around zero. The
open circles mark results before rescattering, the black circles
after rescattering. $N/10$ denotes nucleon multiplicity divided by
10.} \label{fig8}
\end{center}
\end{figure}

\begin{figure}[h!]
\begin{center}
\includegraphics[height=8cm]{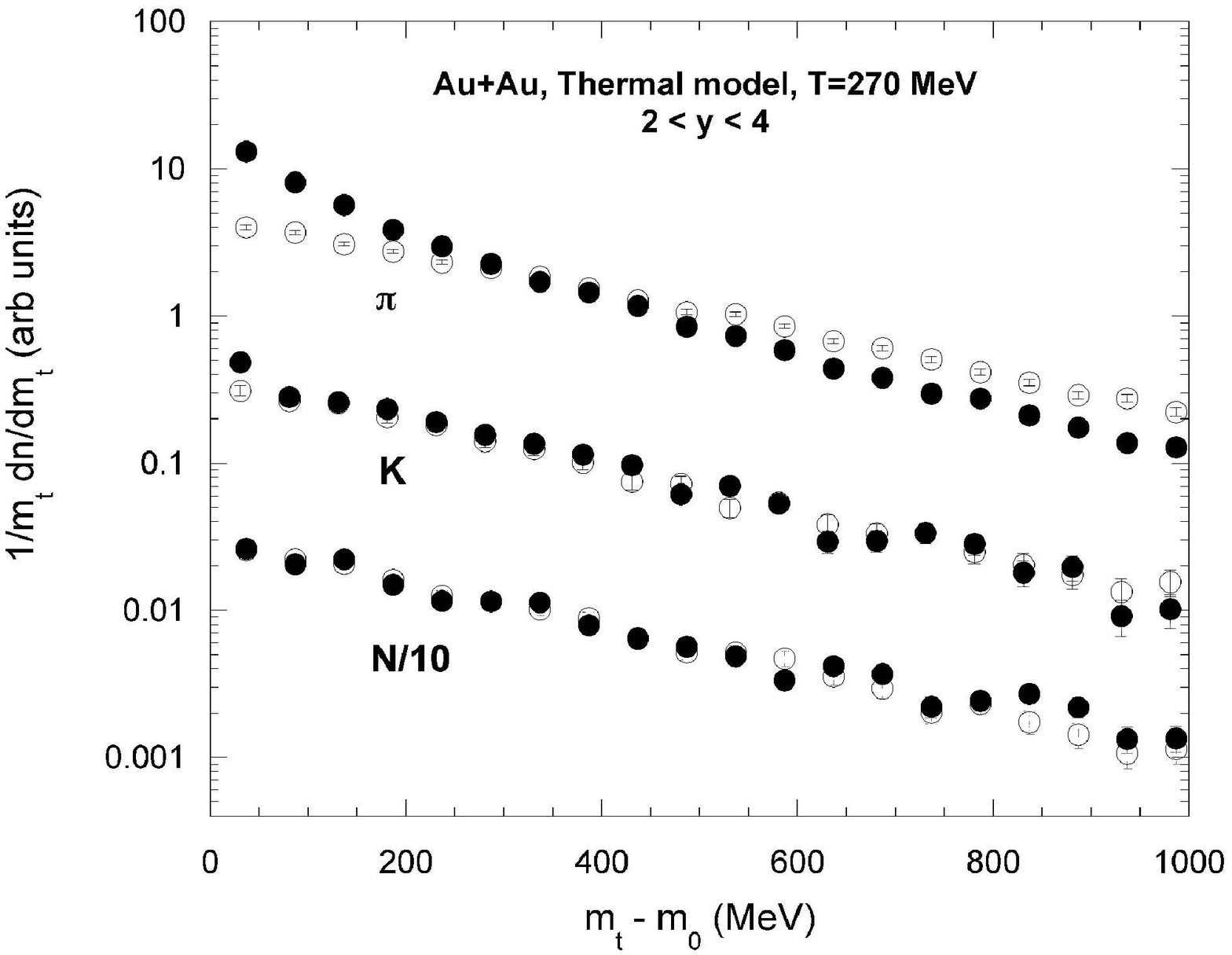} \caption{Transverse mass
spectra for $T=270$ MeV collected for rapidities around three. The
open circles mark results before rescattering, the black circles
after rescattering and $N/10$ denotes nucleon multiplicity divided
by 10.} \label{fig9}
\end{center}
\end{figure}

Figures \ref{fig8} and \ref{fig9} show the $m_{t}$-spectra before
and after rescattering at $T=270$ MeV for rapidity zero and for
rapidities near 3, respectively. The  rescattering produces a
steeper fall off with $m_{t}$ for pions at both rapidity zero and
3.35, the rescattering, so to speak, cools the pions. Kaons are
influenced by the rescattering in a similar way, but to a smaller
degree, while the nucleon spectrum becomes less steep in its fall
off with $m_{t}$ at midrapidity, the nucleons get heated by the
collisions with pions. At $y=3$ the nucleon spectrum is not changed
by the rescattering. The inverse slopes are quantified in Table II,
with the help of exponential fits to the spectra of Figures
\ref{fig8} and \ref{fig9}, where the fitting ranges correspond to
the $m_{t}$ ranges in the figures for each spectrum. The inverse
slopes change with rapidity in a distinct way for each particle
type, reflecting the relative weighting of the contributions from
the middle and forward-backward source-centers. The inverse slopes
also increase markedly with mass of the particle, a feature normally
taken to be indicative of flow, something that has not been
introduced explicitly in the model.

\newpage

\begin{center}
\begin{table}
\caption{Inverse slope parameters (MeV) from the thermal model for
T=270 MeV.}
\begin{tabular}{cclcclcc} \hline
particle & y& after & slope & error & before & slope & error  \\
\hline
$\pi$    & 0          &         & 195 &          1      &           & 351         &   3      \\
K         &             &        & 267 &          5      &            & 300        &   6       \\
N         &             &        & 449  &        13      &
&  301         &  6      \\ \hline
$\pi$    & 3          &        &  195 &         2       &           & 326         &  4       \\
K         &             &        & 259  &         7      &            & 289        &  9        \\
N         &             &        & 313   &        9      &
&  293        &   8       \\ \hline
\end{tabular}
\end{table}
\end{center}

\subsection{Comparison with experimental data}
The rescattering, as discussed above, has little influence on the
model predictions for $dn/dy$, just as the increase in temperature
from 200 to 270 MeV has. Therefore the agreement with data is
similar to what was shown in Figure \ref{fig1}. The figure shows
protons and antiprotons separately and the comparison after
rescattering should rather be with nucleons; a comparison of proton
plus antiproton data, however does not change the quality of the
agreement.

The model predictions for protons and antiprotons at $T=200$ MeV
have $N_{\bar{p}0}/N_{p0}$=0.94 (see Table I), where a value of 1.0
would indicate a midrapidity source center with equal number of
protons and antiprotons, i.e. a baryon chemical potential of zero.
This follows because the F-functions in Eq. (3) are identical for
$p$ and $\bar{p}$.  A fit with an N-ratio of 1 could not be enforced
with the present source-center geometry, as it would require
negative forward and backward antiproton sources.
\begin{figure}[h!]
\begin{center}
\includegraphics[height=8cm]{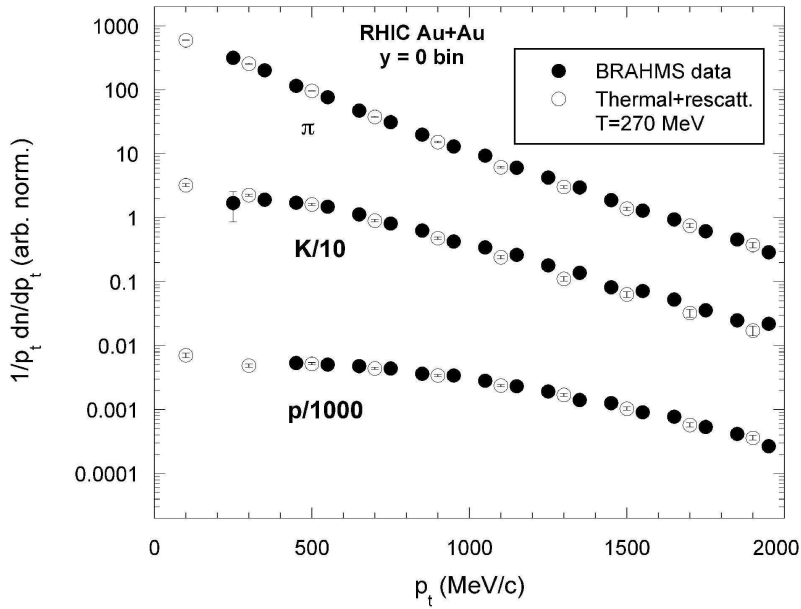} \caption{Comparison of
transverse momentum spectra for $T=270$ MeV and data from BRAHMS at
midrapidity. The data are denoted by open circles and are from
references \cite{Peter 2003} and \cite{Djamel 2004}.  The K-meson
multiplicities have been divided by 10 and the proton data by 1000.
The model results are for nucleons rather than protons.}
\label{fig10}
\end{center}
\end{figure}

\begin{figure}[h!]
\begin{center}
\includegraphics[height=8cm]{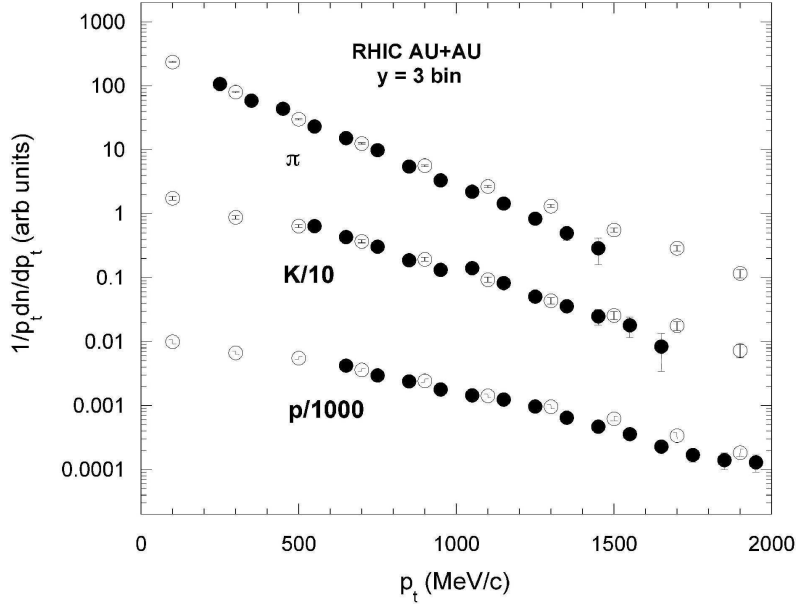} \caption{Comparison of
transverse momentum spectra for $T=270$ MeV and data from BRAHMS at
rapidities about three. The data are denoted by open circles and are
from references \cite{Peter 2003} and \cite{Djamel 2004}.  The
K-meson multiplicities have been divided by 10 and the proton data
by 1000. The model results are for nucleons rather than protons.}
\label{fig11}
\end{center}
\end{figure}
The comparison between data and model for the $p_{t}$-spectra is
shown in Figures \ref{fig10} and \ref{fig11} at 270 Mev. For $y=0$
(Figure \ref{fig10}) the agreement between model and data is good
and at $y=3$ (Figure \ref{fig11}) the agreement is reasonable. Thus
the model can reproduce the data quite well at $T=270$ Mev at both
rapidity intervals. It may be noted that the experimental $p_{t}$
spectra for protons and antiprotons exhibit nearly identical slopes
\cite{Peter 2003} so the comparison of nucleon spectra from the
model calculations to proton data is  valid.

\section{Results from the partonic model with rescattering}
\subsection{Comparison of results with and without rescattering}
\begin{figure}[h!]
\begin{center}
\includegraphics[height=8cm]{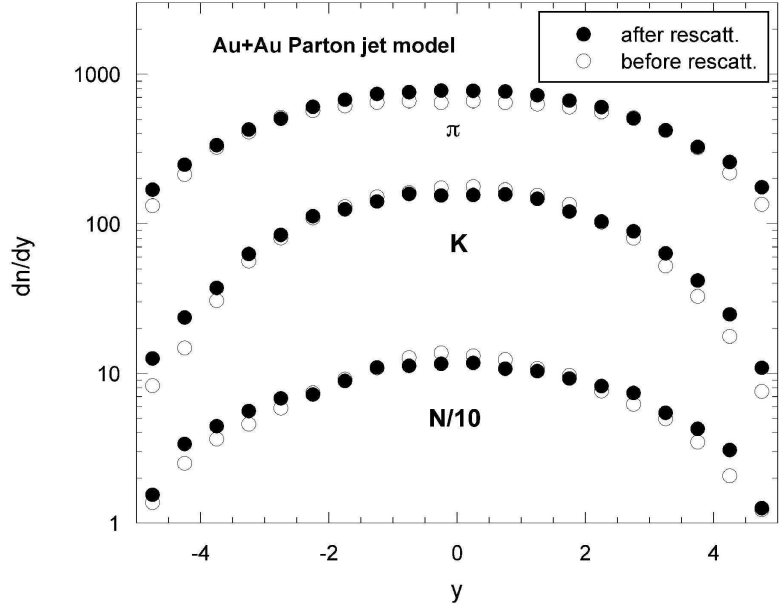} \caption{Rapidity
density $dn/dy$ as plotted versus center of mass rapidity. The
results before rescattering are marked with open symbols, those
after with black circles. The notation $N/10$ stands for nucleon
multiplicity divided by 10.} \label{fig12}
\end{center}
\end{figure}

Figure \ref{fig12} shows the $dn/dy$ distributions for $\pi$, $K$
and nucleons from the partonic model (open circles) and after
rescattering (black circles). The data cover 20 events as for the
thermal model results. In agreement with the thermal model results,
the rescattering changes the distributions by very little, they are
slightly broadened. The $m_{t}$ spectra however are markedly
influenced; the rescattering makes the $\pi$ spectra steeper and
flattens the $K$ and nucleon spectra, in particular at mid-rapidity.
The effect for nucleons at mid-rapidity is somewhat stronger than
for the thermal model, demonstrated by the inverse slopes obtained
from exponential fits quoted in Table III for the partonic model and
in Table II (above) for the thermal model. The changes are in the
same directions as for the thermal model calculations.

\newpage
\begin{center}
\begin{table}
\caption{Inverse slope parameters (MeV) from the partonic jet
model.}
\begin{tabular}{cclcclcc} \hline
particle & y& after & slope & error & before & slope & error  \\
\hline
$\pi$    & 0          &         & 165 &          1      &           & 299         &   2      \\
K         &             &        & 230 &          3      &            & 168        &   2       \\
N         &             &        & 358  &        8      &
&   167       &  2      \\ \hline
$\pi$    & 3          &        &  143 &         1       &           & 196         &  1       \\
K         &             &        & 166  &         3      &            & 141        &  3        \\
N         &             &        & 201   &        4      &
&  150        &   3       \\ \hline
\end{tabular}
\end{table}
\end{center}
The changes for nucleons caused by the rescattering are indeed quite
dramatic, as is the cooling of the pions.

\subsection{Comparison to data}
\begin{figure}[h!]
\begin{center}
\includegraphics[height=6cm]{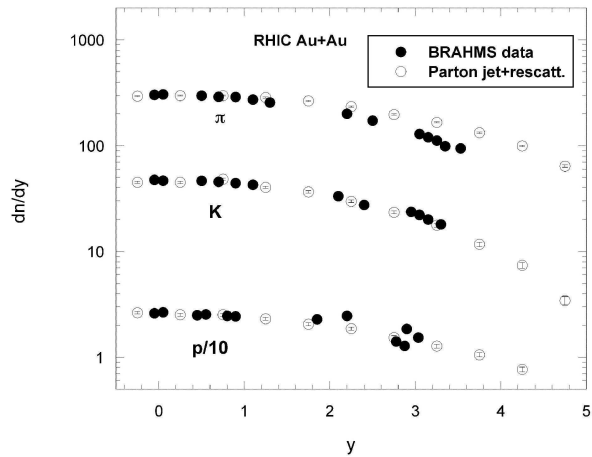} \caption{Rapidity
densities versus center of mass rapidity. The data from
references\cite{Peter 2003} and \cite{Djamel 2004} are plotted with
black circles, while the results from the partonic model (after
rescattering) are shown as open symbols.} \label{fig13}
\end{center}
\end{figure}

\begin{figure}[h!]
\begin{center}
\includegraphics[height=6cm]{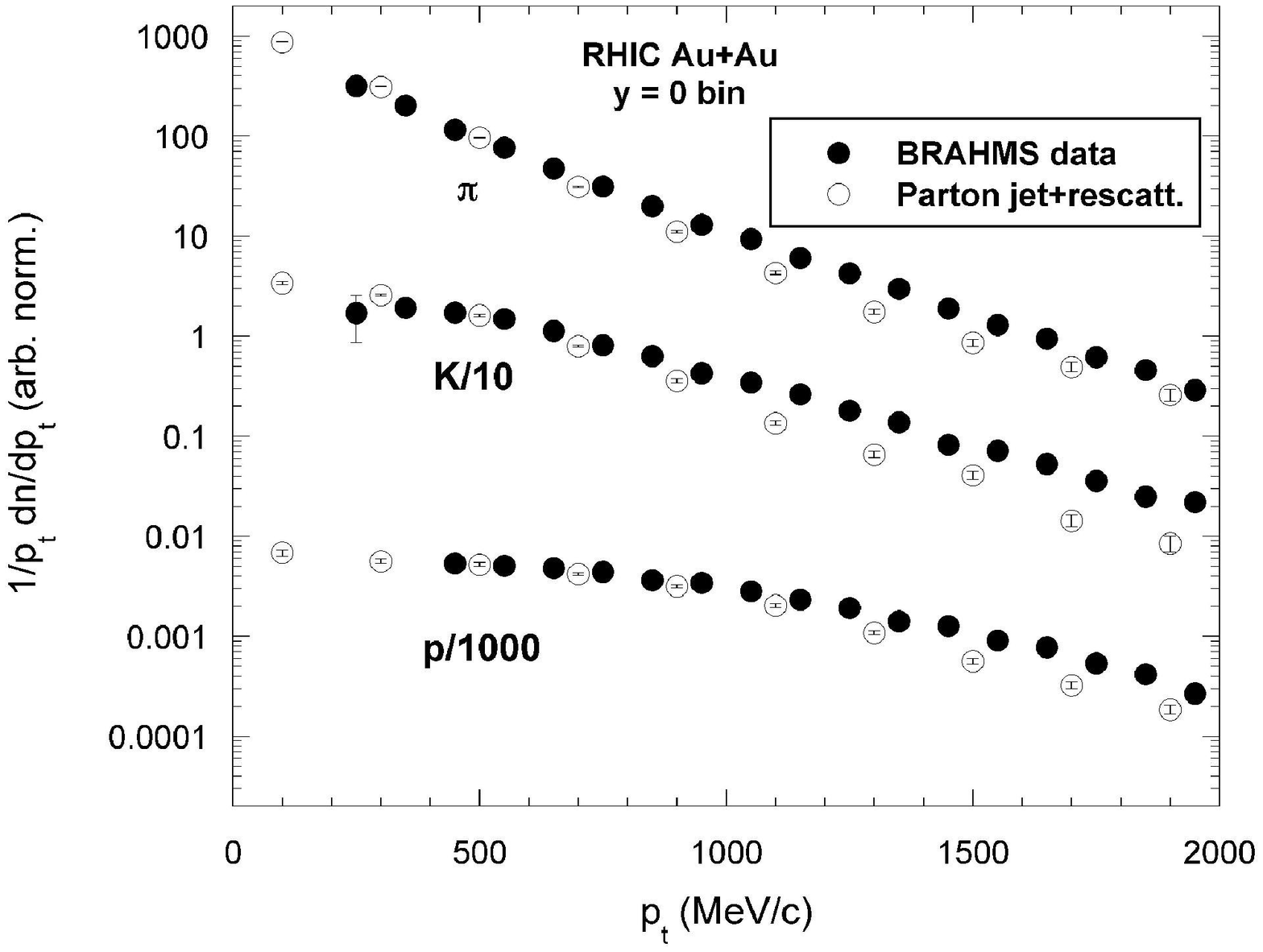} \caption{Transverse
momentum spectra at midrapidity, a comparison between data and
model. The data are denoted by open circles and are from references
\cite{Peter 2003} and \cite{Djamel 2004}.  The K-meson
multiplicities have been divided by 10 and the proton data by 1000.
The model results are for nucleons rather than protons.}
\label{fig14}
\end{center}
\end{figure}

\begin{figure}[h!]
\begin{center}
\includegraphics[height=6cm]{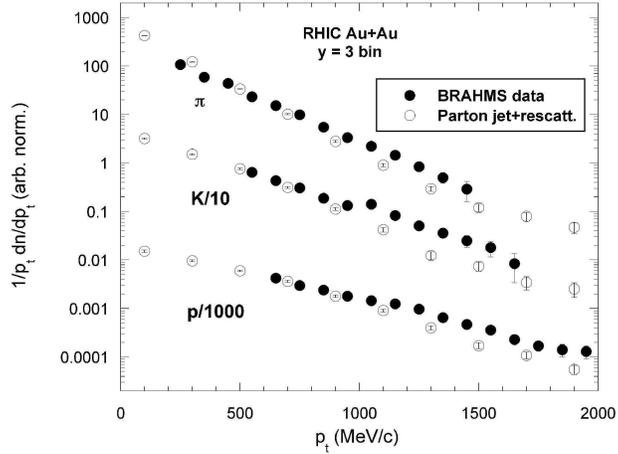} \caption{Transverse
momentum spectra near $y=3$, a comparison between data and model.
The data are denoted by open circles and are from references
\cite{Peter 2003} and \cite{Djamel 2004}.  The K-meson
multiplicities have been divided by 10 and the proton data by 1000.
The model results are for nucleons rather than protons.}
\label{fig15}
\end{center}
\end{figure}

The $dn/dy$ distributions after rescattering are compared to the
data for $\pi$, $K$ and nucleons in Figure \ref{fig13}, where the
data are for protons rather than for nucleons. The agreement between
model predictions and data are reasonable, but not as good as for
the thermal case in Figure \ref{fig1}; it should be remarked,
though, that in the thermal case there was a parameter adjustment
for each particle species. The $p_{t}$ distributions are compared to
data in Figures \ref{fig14} and \ref{fig15}, where the agreement is
very good both at midrapidity and near $y=3$. Inverse slopes vary
with mass (see also Table III) in a way expected for flow, again
without flow appearing explicitly in the calculations. In both
cases, $dn/dy$ and transverse spectra, the data are shown for
protons rather than nucleons ($p+\bar{p}$), but the conclusions
drawn are not affected.

\section {Discussion}
The two event generating models used here are very different and
rather schematic. The thermal-like model has no dynamical features
and therefore little predictive power, the fitting procedure
described in Subsection 2.3 will have to be repeated at each
incident energy, and the direct agreement with experiment without
rescattering regarding the transverse spectra is rather  poor. The
partonic model is in principle a dynamic model with predictive
power, however the pion coalescence mechanism is  certainly {\em ad
hoc} and may have to be adjusted at each incident energy.  Also here
the predicted transverse spectra agree rather poorly with experiment
without rescattering. The main goal in this work has been to study
the effects of hadronic rescattering on the  hadrons produced by the
two different models of the initial stage of the collision, and
although the two models are very different, i.e. thermal-like
hadrons vs. parton jets, after rescattering they both give similar
hadronic rapidity, $m_t$, and $p_t$ distributions, which agree
reasonably well with RHIC experiments. Although rescattering has
only a weak effect on the rapidity distributions for either model,
it is seen to strongly affect the $m_t$ (and $p_t$) distributions.
For the thermal model with $T=270$ MeV, see Table II, the initial
slope parameters are similar for the three particle  species
implying no initial radial flow from the model, whereas after
rescattering the slope parameter increases significantly for
increasing particle mass, and a radial flow effect like the
experiments is seen. The slope parameters from the parton model also
show radial flow and agree with experiments after rescattering, but
as seen in Table III, before rescattering the slope parameters
actually decrease significantly with increasing particle mass
indicating a sort of ``anti-radial-flow" effect. Thus even though
the two models strongly disagree in the $m_t$ distributions they
directly produce for the three particle species, rescattering
effects are able to sufficiently wash out these differences such
that after rescattering the particle distributions are now in
essential agreement. We conclude from this that features seen in
$m_t$ (and $p_t$) distributions before rescattering are mostly due
to the overall temperature  scale of the initial stage and
rescattering effects are not very sensitive to the details of the
initial stage model used. It is however remarkable that the
rescattering changes the spectra in the same fashion as found in an
earlier publication \cite{Humanic:1998a} where a third event
generator model was used: ``cooling" of the pion spectra and
``heating" of the nucleon spectra with the kaons in between.
\begin{figure}[h!]
\begin{center}
\includegraphics[height=7cm]{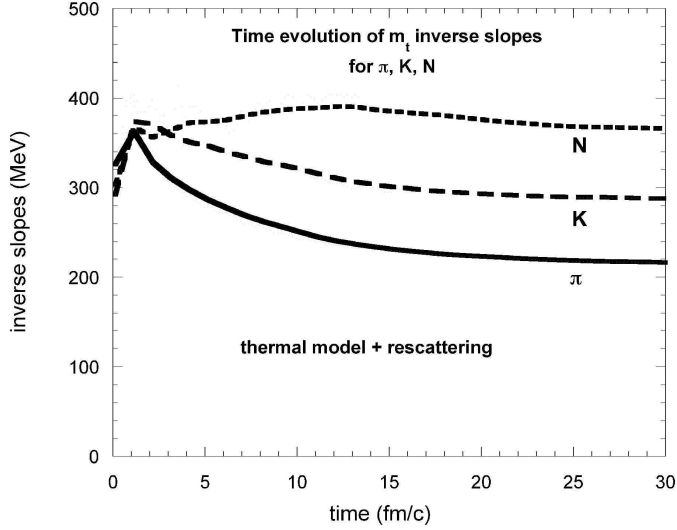} \caption{Inverse slopes
with rescattering for the thermal model plotted versus time. The
inverse slopes were found from exponential fits to the $m_{t}$
spectra at the the various times. See also the text.} \label{fig16}
\end{center}
\end{figure}

\begin{figure}[h!]
\begin{center}
\includegraphics[height=7cm]{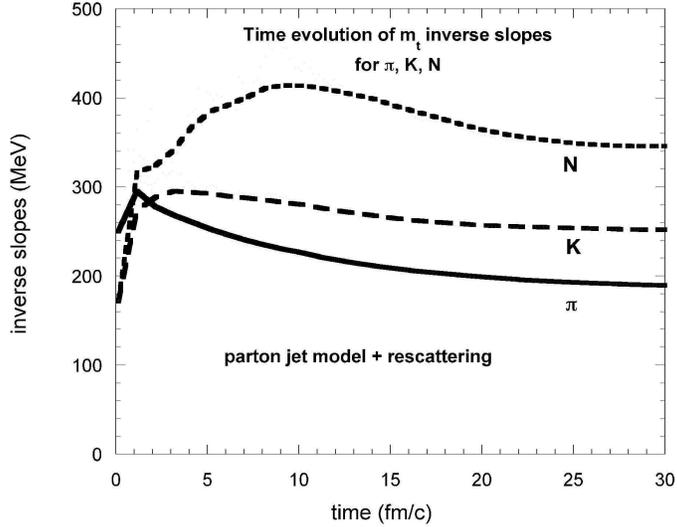} \caption{Inverse slopes
with rescattering for the partonic model plotted versus time.
Notation and definitions are as in the previous figure.}
\label{fig17}
\end{center}
\end{figure}
The two last figures, Figures \ref{fig16} and \ref{fig17}, show the
inverse slopes for $m_{t}$ spectra from the thermal-like and the
partonic model both with rescattering, respectively, plotted versus
time, $t$. The density of particles builds up as $t$ increases from
zero until the volume increase with time overtakes the formation of
new particles and the density starts to decrease, which happens at
about $t=4$ fm/c. At large $t$ a steady state is reached for the
inverse slopes and the pattern of increasing inverse slope with
increasing particle mass, seen from Tables II and III has become
evident. The changes at small times are very fast and the pushing of
the faster pions on the other particles is clear while the pions
themselves loose momentum. The first ten fm/c are very important for
the development of the final slope pattern. The inverse slopes in
the two figures do not agree quantitatively with the numbers in
Tables II and III, at low $t$ because new particles enter fast as
their formation times are reached, while in the tables the spectra
are for all particles at freeze out, and moreover because rapidity
ranges and $m_{t}$ ranges for the exponential fittings are different
between tables and Figures \ref{fig16} and \ref{fig17}. The
conclusion is that to the extent the model approaches used here
reflect in some reasonable way what happens in the real heavy ion
collisions in the laboratory, one should evidently not draw strong
conclusions from the hadronic $y$ and $m_{t}$ spectra neither
regarding the presence of flow nor regarding the initial collision
conditions.

It would be interesting to look at other hadronic observables  such
as elliptic flow and HBT interferometry using these two models to
see if such observables can be used to discriminate between initial
conditions.

\begin{acknowledgments}
The authors wish to acknowledge financial support from the U.S.
National Science Foundation under grant PHY-0355007 and from the
Danish SNF for travel expenses. We would in particular thank Tracy
L. Smith for expert systems management at the Ohio end of the
collaboration.
\end{acknowledgments}

\end{document}